\documentclass[11pt,fleqn]{article}
\usepackage{amssymb}

\usepackage{amsmath}

\usepackage{amsmath,amssymb,amsthm}

\begin{document}

\begin{center}

{\Large \textbf{Reduction of non-linear d'Alembert equations to
two-dimensional equations}}

\vskip 15pt {\large \textbf{Irina YEHORCHENKO}}

\vskip 6pt {Institute of Mathematics of NAS Ukraine, 3
Tereshchenkivs'ka Str., 
\\ 01601 Kyiv-4, Ukraine} \\E-mail:
iyegorch@imath.kiev.ua
\end{center}

\begin{abstract}
We study conditions of reduction of the multidimensional wave
equation $ \Box u = F(u)$ - a system of the d'Alembert and
Hamilton equations: $y_\mu y_\mu = r(y,z); y_\mu z_\mu = q(y,z);
z_\mu z_\mu = s(y,z);$ $ \Box y = R(y,z); \Box z = S(y,z).$ We
prove necessary conditions for compatibility of such system of the
reduction conditions. Possible types of the reduced equations
represent interesting classes of two-dimensional pa\-ra\-bo\-lic,
hyperbolic and elliptic equations. Ansatzes and methods used for
reduction of the d'Alembert ($n$-dimensional wave) equation can be
also used for arbitrary Poincar\'e-invariant equations. This
seemingly simple and partial problem involves many important
aspects in the studies of the PDE. The paper was published in Proceedings 
of the 4th Workshop "Group Analysis of Differential 
Equations \& Integrable Systems", 2009, p. 243-253.\end{abstract}

\section{Introduction}

We study conditions of reduction of the multidimensional wave
equation
\begin{gather}\label{yehorchenko:nl wave}
\Box u=F(u),\\
\Box\equiv \partial^2_{x_0}-\partial^2_{x_1}-\cdots
-\partial^2_{x_n}, \quad u=u(x_0, x_1, \ldots, x_n)\nonumber
\end{gather}
by means of the ansatz with two new independent variables
\cite{yehorchenko:WFans,yehorchenko:GrunlandHarnadWinternitz94}
\begin{equation} \label{yehorchenko:ansatz2}
u=\varphi(y,z),
\end{equation} where $y$, $z$ are new variables.
Henceforth $n$ is the number of independent space variables in the
initial d'Alembert equation.

These conditions are a~system of the d'Alembert and Hamilton
equations:
\begin{gather} \label{yehorchenko:comp conditions}
y_\mu y_{\mu}=r(y,z), \quad y_{\mu} z_{\mu}=q(y,z), \quad
z_{\mu}z_{\mu}=s(y,z), \\
\Box y=R(y, z), \quad \Box z=S(y, z) \nonumber.
\end{gather}

We prove necessary conditions for compatibility of such system of
the reduction conditions. This paper is a development of research
started jointly with W.I. Fushchych in 1990s
\cite{yehorchenko:wave-reduction}, and we present some new results
and ideas.

Possib\-le types of the reduced equations represent interesting
classes of two-dimensional equations - parabolic, hyperbolic 
and elliptic. Ansatzes and methods used for reduction of the
d'Alembert ($n$-dimensional wave) equation can be also used for
arbitrary Poincar\'e-invariant equations. This seemingly simple
and partial problem involves many important aspects in the studies
of the PDE.

Classes of exact solutions of non-linear equations having
respective symmetry properties can be constructed by means of
symmetry reduction of these equations to equations with smaller
number of independent variables or to ordinary differential
equations (for the algorithms see the books
\cite{yehorchenko:OvsR} - \cite{yehorchenko:BlumanKumeiBook}).

Reductions and solutions of equation (\ref{yehorchenko:nl wave})
by means of symmetry reduction or ansatzes were considered in
numerous papers \cite{yehorchenko:Tajiri84}-
\cite{yehorchenko:FBarMosk}. See \cite{yehorchenko:F95} for a
review of results related to reduction of a number of wave
equations. In the paper~\cite{yehorchenko:FBar96} an alternative
was proposed for the method of application of ansatzes for
equation~(\ref{yehorchenko:nl wave}) with a degree nonlinearity.

The method of symmetry reduction does not give exhaustive
description of all exact solutions for an~equation, so other
methods for construction of exact solutions may be expedient.

A so-called ``direct method'' for search of exact solutions of
nonlinear partial differential equations (giving wider classes of
solutions than the symmetry reduction) was proposed by P.~Clarkson
and M.~Kruskal~\cite{yehorchenko:Clarkson} (see
also~\cite{yehorchenko:Clarkson Mansfield CS NLWE,
yehorchenko:Olver94} and the papers cited therein). It is easy to
see that this method for majority of equations results in
considerable difficulties as it requires investigation of
compatibility and solution of cumbersome reduction conditions of
the initial equation. These reduction conditions are much more
difficult for investigation and solution in the case of equations
containing second and/or higher derivatives for all independent
variables, and for multidimensional equations - e.g. in the
situation of the nonlinear wave equations.

The direct method, if  applied ``completely'' (with full solution
of compatibility conditions), is exhaustive to some extent - it
allows obtaining all reductions of the original equations that may
be obtained from $Q$-conditional symmetry (see more comments on
symmetry in Section 4).

In this paper we were not able to achieve such complete
application of the direct reduction to
equation~(\ref{yehorchenko:nl wave}) - the presented results are
only a step to such application. To do that it is necessary to
find a general solution of the reduction conditions.

Direct reduction with utilisation of ansatzes or exhaustive
description of conditional symmetries (even $Q$-conditional
symmetries) cannot be regarded as algorithmic to the same extent
as the standard symmetry reduction. Majority of papers on
application of the direct method are devoted to evolution
equations or other equations that contain variables of the order
not higher than one for at least one of the independent variables,
with not more than three independent variables. In such cases
solution of the reduction conditions is relatively simple.

We consider general reduction conditions of
equation~(\ref{yehorchenko:nl wave}) by means of a general ansatz
with two new independent variables. We found necessary
compatibility conditions for the respective reduction conditions
-- we developed the conditions found
in~\cite{yehorchenko:wave-reduction}. We also describe respective
possible forms of the reduced equations. Thus we proved that the
reduced equations may have only a particular form.

A similar problem was considered by previous authors for an ansatz
with one independent variable
\begin{equation} \label{yehorchenko:ansatz1}
u=\varphi(y),
\end{equation} where $y$ is a new independent variable.

Compatibility analysis of the d'Alembert--Hamilton system
\begin{gather} \label{yehorchenko:dA-H1}
\Box u=F(u), \quad u_{\mu} u_{\mu}=f(u)
\end{gather}
in the three-dimensional space was done in
\cite{yehorchenko:Collins1}. For more detailed review of
investigation and solutions of this system
 see~\cite{yehorchenko:FZhR-UMZh,yehorchenko:FZh-book}.

The compatibility condition of the
system~(\ref{yehorchenko:dA-H1}) for $f(u)~=~0$ was found in the
paper~\cite{yehorchenko:CieciuraGrundland}.

Complete investigation of compatibility of overdetermined systems
of differential equations with fixed number of independent
variables may be done by means of Cartan's
algorithm~\cite{yehorchenko:Cartan}, however, it is very difficult
for practical application even in the case of three independent
variables, and not applicable for arbitrary number of independent
variables. For this reason some ad hoc techniques for such cases
should be used even for search of necessary compatibility
conditions.

It is evident that the d'Alembert--Hamilton
system~(\ref{yehorchenko:dA-H1}) may be reduced by local
transformations to the form
\begin{gather} \label{yehorchenko:dA-H1-lambda}
\Box u=F(u), \quad u_{\mu} u_{\mu}=\lambda, \quad \lambda=0,\pm1.
\end{gather}

Necessary compatibility conditions of the
system~(\ref{yehorchenko:dA-H1-lambda}) for four independent
variables were studied in \cite{{yehorchenko:FZh PLA}} (see
also~\cite{yehorchenko:FZh-book}). The necessary compatibility
conditions for the system (\ref{yehorchenko:dA-H1-lambda}) for
arbitrary number of independent variables were found in
\cite{yehorchenko:FZhY}:

{\bf Statement 1}. {\it For the
system~\eqref{yehorchenko:dA-H1-lambda} ($n$ is arbitrary) to be
compatible it is necessary that the function $F$ have the
following form:}
\[
F=\frac{\lambda \partial_u \Phi}{\Phi}, \quad
\partial_u^{n+1}\Phi=0.
\]

W.I.~Fushchych, R.Z.~Zhdanov and I.V.~Revenko
\cite{{yehorchenko:FZhR-UMZh},{yehorchenko:FZhR-preprint},{yehorchenko:FZhR-JMPh}}
found a general solution of the system
(\ref{yehorchenko:dA-H1-lambda}) for three space variables (that
is four independent variables), as well as necessary and
sufficient compatibility conditions for this
system~\cite{yehorchenko:FZhR-preprint}:

{\bf Statement 2}. {\it For the
system~\eqref{yehorchenko:dA-H1-lambda} $(u=u(x_0,x_1,x_2,x_3))$
to be compatible it is necessary and sufficient that the function
$F$ have the following form:}
\[
F=\frac{\lambda}{N(u+C)}, \quad N=0,1,2,3.
\]

The results presented in this paper may be regarded as development
of the above Statements.

Reduction of equation~(\ref{yehorchenko:nl wave}) by means of the
ansatz~(\ref{yehorchenko:ansatz2}) was considered
in~\cite{yehorchenko:FZhR-JMPh} for a special case (when the
second independent variable enters the reduced equation only as a
parameter), all respective ansatzes for the case of four
independent variables were described, and the respective solutions
were found. Some solutions of such type for arbitrary $n$ were
also considered in~\cite{yehorchenko:BarYur}.

In \cite{yehorchenko:zhdanov-panchakBoxu} reduction of the
nonlinear d'Alembert equation by means of ansatz
$u=\phi(\omega_1,\omega_2,\omega_3)$ was considered for the case
$\Box\omega_1=0$, $\omega_{1\mu}\omega_{1\mu}=0$ (that is
$\omega_1$ entered the reduced equation only as a parameter). The
respective compatibility conditions were studied and new (non-Lie)
exact solutions were found. Note that this case does not include
completely the case considered here - the case of the ansatz with
two new independent variables.

\section{Necessary compatibility conditions of the system of the
d'Alem\-bert--Hamilton equations for two functions or for a
complex-valued function}

Reduction of multidimensional equations to two-dimensional ones
may be interesting as solutions of two-dimensional partial
differential equations, including non-linear ones, may be
investigated more comprehensively than solutions of
multidimensional equations, and such two-dimensional equations may
have more interesting properties than ordinary differential
equations. Two-dimensional reduced equations also may have
interesting properties with respect to conditional symmetry.

Substitution of ansatz~(\ref{yehorchenko:ansatz2}) into the
equation~(\ref{yehorchenko:nl wave}) leads to the following
equation:
\begin{gather} \label{yehorchenko:reduction}
\varphi_{yy}y_{\mu}y_{\mu}+2\varphi_{yz}z_{\mu}y_{\mu}+
\varphi_{zz}z_{\mu}z_{\mu}+\varphi_y \Box y+\varphi_z \Box
z=F(\varphi)\\
\left(y_{\mu}=\frac{\partial y}{\partial x_{\mu}}, \ \
\varphi_y=\frac{\partial \varphi}{\partial y}\right),\nonumber
\end{gather}
whence we get a system of equations:
\begin{gather} \label{yehorchenko:comp conditions}
y_\mu y_{\mu}=r(y,z), \quad y_{\mu} z_{\mu}=q(y,z), \quad
z_{\mu}z_{\mu}=s(y,z), \\
\Box y=R(y, z), \quad \Box z=S(y, z) \nonumber.
\end{gather}

System (\ref{yehorchenko:comp conditions}) is a reduction
condition for the multidimensional wave
equation~(\ref{yehorchenko:nl wave}) to the two-dimensional
equation~(\ref{yehorchenko:reduction}) by means of
ansatz~(\ref{yehorchenko:ansatz2}).

The system of equations (\ref{yehorchenko:comp conditions}),
depending on the sign of the expression~$rs-q^2$, may be reduced
by local transformations to one of the following types:

1) elliptic case: $rs-q^2>0$, $v=v(y,z)$~is a complex--valued
function,
\begin{gather}
\Box v=V(v, v^*), \quad \Box v^*=V^*(v, v^*), \nonumber \\
v^*_{\mu}v_{\mu}=h(v, v^*), \quad v_{\mu}v_{\mu}=0,
 \quad v^*_{\mu} v^*_{\mu}=0\label{yehorchenko:ellipt}
\end{gather}
(the reduced equation is of the elliptic type);

2) hyperbolic case: $rs-q^2 < 0$, $v=v(y, z)$, $w=w(y, z)$~are
real functions,
\begin{gather}
\Box v=V(v, w), \quad \Box w = W(v, w), \nonumber\\
w_{\mu}w_{\mu}=h(v, w), \quad v_{\mu}v_{\mu}=0, \quad w_{\mu}
w_{\mu}=0\label{yehorchenko:hyperb}
\end{gather}
(the reduced equation is of the hyperbolic type);

3)  parabolic case: $rs-q^2=0$, $r^2+s^2+q^2\not=0$, $v(y,z)$,
$w(y,z)$~are real functions,
\begin{gather}
\Box v=V(v,w), \quad \Box w = W(v,w), \nonumber \\
v_{\mu}w_{\mu}=0, \quad v_{\mu}v_{\mu}=\lambda
 \ (\lambda=\pm 1), \quad w_{\mu} w_{\mu}=0 \label{yehorchenko:parab}
\end{gather}
(if $W\not=0$, then the reduced equation is of the parabolic
type);

4) first-order equations: ($r=s=q=0$), $y \to v$, $z \to w$
\begin{gather}
 v_{\mu}v_{\mu}=w_{\mu} w_{\mu}=v_{\mu}w_{\mu}=0, \nonumber\\
\Box v=V(v, w), \quad \Box w=W(v, w). \label{yehorchenko:1order}
\end{gather}

Let us formulate necessary compatibility conditions for the
systems (\ref{yehorchenko:ellipt})--(\ref{yehorchenko:1order}).

\smallskip

{\bf Theorem 1}. {\it System \eqref{yehorchenko:ellipt} is
compatible if and only if}
\[
V=\frac{h(v,v^*)\partial_{v^*}\Phi}{\Phi}, \quad
\partial_{v^*}\equiv \frac{\partial}{\partial v^*},
\]
{\it where $\Phi$ is an arbitrary function for which the following
condition is satisfied}
\[
(h\partial_{v^*})^{n+1}\Phi=0.
\]

The function $h$ may be represented in the form
$h=\frac{1}{R_{vv^*}}$, where $R$ is an arbitrary sufficiently
smooth function, $R_v$, $R_{v^*}$ are partial derivatives by the
respective variables.

Then the function $\Phi$ may be represented in the form
$\Phi=\sum\limits_{k=0}^{n+1}f_k(v)R_v^k$, where~$f_k(v)$ are
arbitrary functions, and
\[
V=\frac{\sum\limits_{k=1}^{n+1}kf_k(v)R_v^k}{\sum\limits_{k=0}^{n+1}f_k(v)R_v^k}.
\]

The respective reduced equation will have the form
\begin{equation} \label{yehorchenko:ellipt-red1}
h(v,v^*)\left(\phi_{vv^*}+\phi_{v}\frac{\partial_{v^*}\Phi}{\Phi}+\phi_{v^*}\frac{\partial_v\Phi^*}{\Phi^*}\right)=
F(\phi).
\end{equation}

The equation (\ref{yehorchenko:ellipt-red1}) may also be rewritten
as an equation with two real independent variables
($v=\omega+\theta$, $v^*=\omega-\theta$):
\begin{equation} \label{yehorchenko:ellipt-red2}
2\widetilde{h}(\omega,\theta)(\phi_{\omega \omega}+ \phi_{\theta
\theta}) + \Omega(\omega,\theta) \phi_{\omega} +
\Theta(\omega,\theta) \phi_{\theta}= F(\phi).
\end{equation}
We will not adduce here cumbersome expressions for~$\Omega$,
$\Theta$ that may be found from~(\ref{yehorchenko:ellipt-red1}).

\smallskip

{\bf Theorem 2}. {\it System \eqref{yehorchenko:hyperb} is
compatible if and only if}
\[
V=\frac{h(v,w)\partial_{w}\Phi}{\Phi}, \quad
W=\frac{h(v,w)\partial_v\Psi}{\Psi},
\]
{\it where the functions $\Phi$, $\Psi$ are arbitrary functions
for which the following conditions are satisfied}
\[
(h\partial_v)^{n+1}\Psi=0, \quad (h\partial_w)^{n+1}\Phi=0.
\]

The function $h$ may be presented in the
form~$h=\frac{1}{R_{vw}}$, where $R$ is an arbitrary sufficiently
smooth function, $R_v$, $R_w$ are partial derivatives by the
respective variables. Then the functions~$\Phi$, $\Psi$ may be
represented in the form
\[
\Phi=\sum_{k=0}^{n+1}f_k(v)R_v^k, \quad
\Psi=\sum_{k=0}^{n+1}g_k(w)R_w^k,
\]
where $f_k(v)$, $g_k(w)$ are arbitrary functions, ³
\[
V=\frac{\sum\limits_{k=1}^{n+1}kf_k(v)R_v^k}{\sum\limits_{k=0}^{n+1}f_k(v)R_v^k},
\quad
W=\frac{\sum\limits_{k=1}^{n+1}kg_k(w)R_w^k}{\sum\limits_{k=0}^{n+1}g_k(w)R_w^k}.
\]
The respective reduced equation will have the form
\begin{equation} \label{yehorchenko:hyperb-red}
h(v,w)\left(\phi_{vw}+\phi_{v}\frac{\partial_w\Phi}{\Phi}+\phi_{w}\frac{\partial_v\Psi}{\Psi}\right)=
F(\phi).
\end{equation}

{\bf Theorem 3}. {\it System~\eqref{yehorchenko:parab} is
compatible if and only if}
\[
V=\frac{\lambda \partial_v \Phi}{\Phi}, \quad
\partial_v^{n+1}\Phi=0, \quad W\equiv 0.
\]

Equation (\ref{yehorchenko:nl wave}) by means of
ansatz~(\ref{yehorchenko:ansatz2}) cannot be reduced to a
parabolic equation -- in this case one of the variables will enter
the reduced ordinary differential equation of the first order as a
parameter.

Compatibility and solutions of such system for $n=3$ were
considered in~\cite{yehorchenko:FZhR-JMPh}; for this case
necessary and sufficient compatibility conditions, as well as a
general solution, were found.

System (\ref{yehorchenko:1order}) is compatible only in the case
if $V=W\equiv 0$, that is the reduced equation may be only an
algebraic equation~$F(u)$=0. Thus we cannot reduce
equation~(\ref{yehorchenko:nl wave}) by means of
ansatz~(\ref{yehorchenko:ansatz2}) to a first-order equation.

Proof of these theorems is done by means of utilisation of lemmas
similar to those adduced in~\cite{{yehorchenko:FZh
PLA},{yehorchenko:FZhY}}, and of the well-known Hamilton--Cayley
theorem, in accordance to which a matrix is a root of its
characteristic polynomial.

We will present an outline of proof of Theorem 2 for the
hyperbolic case. For other cases the proof is similar.

We will operate with matrices of dimension $(n+1)\times(n+1)$ of
the second variable of the functions $v$ and $w$:
\[
\hat{V}=\{v_{\mu \nu}\},\quad \hat{W}=\{w_{\mu \nu}\}.
\]
With respect to operations with these matrices we utilise
summation arrangements customary for the Minkowsky space:
$v_0=i\partial_{x_0}$, $v_a=-i\partial_{x_a} (a=1,\ldots,n)$,
$v_\mu v_\mu=v_0^2 -v_1^2 -\cdots-v_n^2$.

\smallskip

{\bf Lemma 1}. {\it If the functions $v$ and $w$ are solutions of
the system~\eqref{yehorchenko:hyperb}, then the following
relations are satisfied for them for any $k$}:
\begin{gather*}
{\rm tr}\hat{V}=
\frac{(-1)^k}{(k-1)!}(h(v,w)\partial_{w})^{k+1}V(v,w),
\\
{\rm tr}\hat{W}= \frac{(-1)^k}{(k-1)!}(h(v,w)
\partial_{v})^{k+1}W(v,w).
\end{gather*}

{\bf Lemma 2}. {\it If the functions $v$ and $w$ are solutions of
the system~\eqref{yehorchenko:hyperb}, then $\det \hat{V}=0$,
$\det \hat{W}=0$}.

{\bf Lemma 3}. {\it Let $M_k(\hat{V})$ be the sum of principal
minors of the order $k$ for the matrix $\hat{V}$. If the functions
$v$ and $w$ are solutions of the system
\eqref{yehorchenko:hyperb}, then the following relations are
satisfied for them for any $k$:
\[
M_k(\hat{V})=\frac{(h(v,w)\partial_{w})^k \Phi}{k! \Phi}, \quad
M_k(\hat{W})=\frac{(h(v,w)\partial_v)^k \Psi}{k! \Psi},
\]
where the functions $\Phi$, $\Psi$ satisfy the following
conditions}
\[
(h\partial_v)^{n+1}\Psi=0, \quad (h\partial_w)^{n+1}\Phi=0.
\]

These lemmas may be proved with the method of mathematical
induction similarly to~\cite{yehorchenko:FZhY} with utilisation of
the Hamilton--Cayley theorem ($E$ is a unit matrix of the
dimension $(n+1)\times(n+1)$).
\[
\sum_{k=0}^{n-1}(-1)^k M_k \hat{V}^{n-k}+(-1)^n E \det\hat{V}=0.
\]

It is evident that the statement of Theorem 2 is a direct
consequence of Lemma~3 for $k=1$.

\smallskip

{\bf Note 1}. Equation (\ref{yehorchenko:ellipt}) may be rewritten
for a pair of real functions $\omega={\rm Re}\, v$, $\theta = {\rm
Im}\,v$. Though in this case necessary the respective
compatibility conditions would have extremely cumbersome form.

\smallskip

{\bf Note 2}. Transition from (\ref{yehorchenko:comp conditions})
to (\ref{yehorchenko:ellipt})--(\ref{yehorchenko:1order}) is
convenient only from the point of view of investigation of
compatibility. The sign of the expression $rs-q^2$ may change for
various $y$, $z$, and the transition is being considered only
within the region where this sign is constant.

\smallskip

\section{Examples of solutions of the system of d'Alembert--Hamilton
equations}

Let us adduce explicit solutions of systems of the
type~(\ref{yehorchenko:comp conditions}) and the respective
reduced equations. Parameters $a_{\mu}$, $b_{\mu}$, $c_{\mu}$,
$d_{\mu}$ $(\mu = {0,1,2,3})$ satisfy the conditions:
\begin{gather*}
-a^2=b^2=c^2=d^2=-1 \quad (a^2\equiv a^2_0-a^2_1-\cdots - a^2_3),
\\
ab=ac=ad=bc=bd=cd=0;
\end{gather*}
$y$, $z$~are functions of $x_0$, $x_1$, $x_2$, $x_3$.
\begin{alignat*}{3}
& 1)\quad  && y=ax, \quad z=dx,\quad
\varphi_{yy}-\varphi_{zz}=F(\varphi);&
\\
& 2)\quad && y=ax, \quad  z=\left((bx)^2+(cx)^2+(dx)^2\right)^{1/2},&\\
&&& \varphi_{yy}-\varphi_{zz}-\frac{2}{z}\varphi_z=F(\varphi);&
\end{alignat*}
In this case the reduced equation is a so-called radial wave
equation, the symmetry and solutions of which were investigated
in~\cite{{yehorchenko:AncoLiu},{yehorchenko:YeVconf2003}}.
\begin{alignat*}{3}
& 3)\quad && y=bx+\Phi(ax+dx), \quad z=cx, \quad
-\varphi_{zz}-\varphi_{yy}=F(\varphi);&\\
& 4)\quad && y=\left((bx)^2+(cx^2)\right)^{1/2}\!, \quad z=ax+dx,
\quad \! -\varphi_{yy}-\frac{1}{y}\varphi_{y}=F(\varphi).\!&
\end{alignat*}

\section{Symmetry aspects}
Solutions obtained by the direct reduction are related to symmetry
properties of the equation -- $Q$-conditional symmetry of this
equation
\cite{yehorchenko:FSS,yehorchenko:F87,yehorchenko:FZh-UMZh}
(symmetries of such type are also called non-classical or non-Lie
symmetries
\cite{yehorchenko:Clarkson,yehorchenko:OlverRosenau,yehorchenko:LeviWinternitz}).
For more theoretical background of conditional symmetry and
examples see also \cite{yehorchenko:F95},
\cite{yehorchenko:zhdanov&tsyfra&popovych99},
\cite{yehorchenko:Cicogna-conf03}.

Conditional symmetry and solutions of various non-linear
two-dimensional wave equations that may be regarded as reduced
equations for equation~(\ref{yehorchenko:nl wave}) were considered
in~\cite{yehorchenko:YeVheat}-\cite{yehorchenko:Cicogna conf05}.
It is also possible to see from these papers that symmetry of the
two-dimensional reduced equations is often wider than symmetry of
the initial equation, that is the reduction to two-dimensional
equations allows to find new non-Lie solutions and hidden
symmetries of the initial equation (see e.g.
\cite{yehorchenko:Abraham}).

The Hamilton equation may also be considered, irrespective of the
reduction problem, as an additional condition for the d'Alembert
equation that allows extending the symmetry of this equation. The
symmetry of the system
\[
\Box u=F(u), \quad u_\mu u_\mu~=0
\]
was described in~\cite{yehorchenko:Shulga85}.
In~\cite{yehorchenko:FZhY} a conformal symmetry of the
system~(\ref{yehorchenko:dA-H1}) was found that is a new
conditional symmetry for the d'Alembert equation. Conditional
symmetries of this system were also described
in~\cite{{yehorchenko:FZhR-JMPh},{yehorchenko:zhdanov-panchakBoxu}}.

\section{Conclusions} 
The results of investigation of
compatibility and solutions of the systems
(\ref{yehorchenko:ellipt})--(\ref{yehorchenko:1order}) may be
utilised for investigation and search of solutions also of other
Poincar\'e--invariant wave equations, beside the d'Alembert
equation, e.g. Dirac equation, Maxwell equations and equations for
the vector potential.

Thus, in the present paper we found
\begin{enumerate}\vspace{-1mm}\itemsep=-1pt
\item[1)] necessary compatibility conditions for the system of the
d'Alembert--Hamilton equations for two dependent functions, that
is reduction conditions of the non-linear multidimensional
d'Alembert equation by means of ansatz~(\ref{yehorchenko:ansatz2})
to a two-dimensional equation; such compatibility conditions for
equations of arbitrary dimensions cannot be found by means of the
standard procedure.

\item[2)] possible types of the two-dimensional reduced equations
that may be obtained from equation~(\ref{yehorchenko:nl wave}) by
means of ansatz~(\ref{yehorchenko:ansatz2}).\vspace{-1mm}
\end{enumerate}

The found reduction conditions and types of ansatzes may be also
used for arbitrary Poincar\'e--invariant multidimensional
equation. In \cite{yehorchenko:F Ye DifInvs} the general form of
the scalar Poincar\'e--invariant multidimensional equations were
described; it is easy to prove that by means of
ansatz~(\ref{yehorchenko:ansatz2}) it is possible to reduce all
these equations to PDE in two independent variables.

\section{Further Research}

\begin{enumerate}\vspace{-1mm}\itemsep=-1pt

\item Study of Lie and conditional symmetry of the system of the
reduction conditions (symmetry of the system of the d'Alembert
equations for the complex function was investigated
in~\cite{{yehorchenko:FYe dA*89}}).

\item Investigation of Lie and conditional symmetry of the reduced
equations. Finding exact solutions of the reduced equations.

\item Relation of the equivalence group of the class of the
reduced equations with symmetry of the initial equation.

\item Group classification of the reduced equations.

\item Finding of sufficient compatibility conditions and of
 a general solution of the compatibility
conditions for lower dimensions~($n=2,3$).

\item Finding and investigation of compatibility conditions and
classes of the reduced equations for other types of equations, in
particular, for Poincar\'e--invariant scalar equations.
\end{enumerate}

\subsection*{Acknowledgements}

I would like to thank the organisers of the Workshop on Group
Analysis of Differential Equations and Integrable Systems for the
wonderful conference and support of my participation.

\end{document}